\documentclass[submission, Proceedings]{SciPost}
\hypersetup{
    colorlinks,
    linkcolor={red!50!black},
    citecolor={blue!50!black},
    urlcolor={blue!80!black}
}

\usepackage[bitstream-charter]{mathdesign}
\DeclareSymbolFont{usualmathcal}{OMS}{cmsy}{m}{n}
\DeclareSymbolFontAlphabet{\mathcal}{usualmathcal}

\usepackage{siunitx}
\usepackage{subcaption}

\renewcommand{\eqref}[1]{(\ref{#1})}

\newcommand{\figref}[1]{Figure~\ref{#1}}

\numberwithin{equation}{section}
\numberwithin{figure}{section}
\numberwithin{table}{section}

\begin{document}

% For convenience during refereeing (optional),
% you can turn on line numbers by uncommenting the next line:
%\linenumbers

% =========================================================================

\begin{center}{\Large \textbf{
Measurement of the transverse polarization of electrons emitted in
neutron decay -- nTRV experiment\\
}}\end{center}

% TODO: write the author list here. Use initials + surname format.
% Separate subsequent authors by a comma, omit comma at the end of the list.
% Mark the corresponding author with a superscript *.
\begin{center}
K. Bodek\textsuperscript{1$\star$} and
A. Kozela\textsuperscript{2}
\end{center}

\begin{center}
{\bf \textsuperscript{1}}
M. Smoluchowski Institute of Physics, Jagiellonian University,
Cracow, Poland
\\
{\bf \textsuperscript{2}}
H. Niewodnicza\'nski Institute of Nuclear Physics, Polish Academy of
Sciences, Cracow, Poland
\\
* kazimierz.bodek@uj.edu.pl
\end{center}

%==================================
\begin{center}
\today
\end{center}

\definecolor{palegray}{gray}{0.95}
\begin{center}
\colorbox{palegray}{
  \begin{tabular}{rr}
  \begin{minipage}{0.05\textwidth}
    \includegraphics[width=24mm]{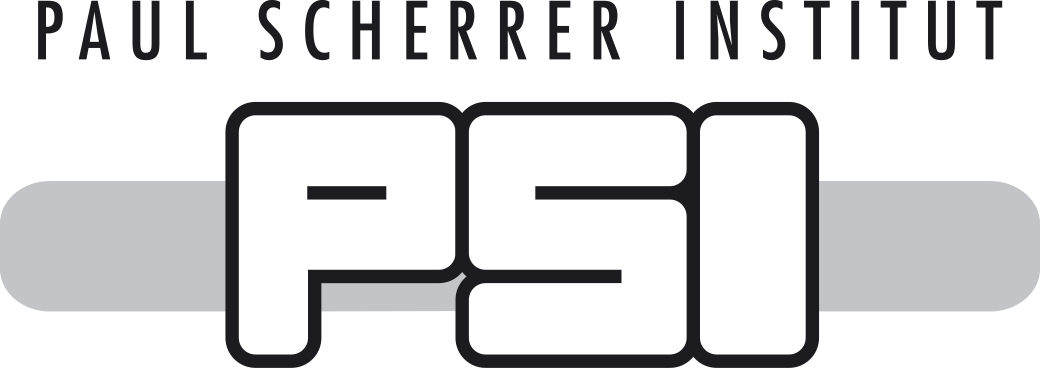}
  \end{minipage}
  &
  \begin{minipage}{0.82\textwidth}
    \begin{center}
    {\it Review of Particle Physics at PSI}\\
    \doi{10.21468/SciPostPhysProc.2}\\
    \end{center}
  \end{minipage}
\end{tabular}
}
\end{center}
%=================================================

\section*{Abstract}
{\bf \boldmath This paper recalls the main achievements of the nTRV experiment which measured two components of the transverse polarization ($\sigma_{T_{1}}$, $\sigma_{T_{2}}$) of
  electrons emitted in the $\beta$-decay of polarized, free neutrons and deduced two correlation coefficients, $R$ and $N$, that are sensitive to physics beyond the Standard Model. The value of time-reversal odd coefficient $R$, 0.004$\pm$0.012$\pm$0.005, significantly
  improved limits on the relative strength of imaginary scalar
  coupling constant in the weak interaction. The value obtained for the time-reversal even
  correlation coefficient $N$, 0.067$\pm$0.011$\pm$0.004, agrees with the Standard Model expectation,
  providing an important sensitivity test of the electron polarimeter. One of the conclusions of this pioneering experiment was that the transverse electron polarization in the neutron $\beta$-decay is worth  more systematic exploring by measurements of yet experimentally not attempted correlation coefficients such as $H$, $L$, $S$, $U$ and $V$. This article presents a brief outlook on that questions.
}

\setcounter{section}{15}
\label{sec:nTRV}

\subsection{Introduction}
\label{nTRV:intro}

Nuclear and neutron beta decay have played a central role in the
development of the weak interaction theory. Among the empirical
foundations of the electroweak sector of the Standard Model (SM), the assumptions of
maximal parity violation, the vector and axial-vector character, and
massless neutrinos are directly linked to nuclear and neutron beta
decay experiments. Beta decay theory was firmly established about six
decades ago and became a part of the SM. It describes the
semi-leptonic and strangeness-conserving processes in the 1-st
particle generation mediated by charged $W$-boson exchange. Despite the neutrino masses have been shown to be finite -- beta
decay experiments with increasing precision still confirm the first
two assumptions. Nevertheless, many open
questions remain such as the origin of parity violation, the hierarchy
of fermion masses, the number of particle generations, the mechanism
of CP violation, and the unexplained large number of parameters of the
theory. A major breakthrough would be a discovery of new CP- or T-violation sources different from the CKM matrix induced mechanism reported for heavier systems in \cite{Christenson1964,Abe2002}. Especially interesting are processes in the systems built of light quarks with vanishingly small contributions of the CKM matrix mechanism such as nuclear beta decay. Therein, experiments with free neutrons play a particularly important role since their interpretation is free of complications connected with nuclear and atomic
structure. In addition, the effects of $p-e$ electromagnetic interaction in the final state, which can mimic T-violation, are small and can be
calculated with a relative precision better than 1\%
\cite{Vogel1983,Ivanov2019,Ivanov2020}.

The nTRV project at PSI, was the first experimental search
for the real and imaginary parts of the scalar and tensor couplings using
the measurement of the transverse polarization of electrons emitted in
the free neutron decay. There are very few measurements of this
observable in general \cite{Danneberg2005,Abe2004}, and only two in
nuclear beta decays. One of them, for the $^{8}$Li system
\cite{Huber2003}, provides the most stringent limit on the tensor coupling
constants of the weak interaction.

According to \cite{Jackson1957},
the decay rate distribution from polarized neutrons as a function of
electron energy ($E$) and momentum (${\bf p}$) is proportional to:
 \begin{equation}\label{Wprob}
  \omega({\bf J}, \mbox{\boldmath$ \hat{\sigma}$}, E, {\bf p}) \varpropto
 \dfrac{}{} 1  + \frac{\langle\bf J\rangle}{J} \cdot
\left( A\frac{{\bf p}}{E} + N {\mbox{\boldmath$ \hat{\sigma}$}} +R\; \frac{{\bf p} \times \mbox{\boldmath$ \hat{\sigma}$}}{E} \right) +\dots
 \end{equation}
where $\frac{\langle\bf J\rangle}{J}$ ($J=|\textbf{J}|$) is the
neutron polarization, \mbox{\boldmath$\hat{\sigma}$} is the unit
vector onto which the electron spin is projected, and $A$ is the beta
decay asymmetry parameter.  $N$ and $R$ are correlation coefficients
which, for neutron decay with usual SM assumptions: $C_{V}\!= \!C'_{V}
\!= \!1$, $C_{A} \!= \!C'_{A}\! = \!\lambda \!=\! -1.276$
\cite{Zyla2020} and allowing for a small admixture of scalar and
tensor couplings $C_{S}$, $C_{T}$, $C'_{S}$, $C'_{T}$, can be
expressed as:
 \begin{eqnarray}\label{nTRV:NR_N}
N &=& \!-0.218\cdot\! \mathrm{Re}(\mathfrak{S}) + 0.335\cdot\! \mathrm{Re}(\mathfrak{T}) - \frac{m}{E}\cdot\! A, \\
R &=& \!-0.218\cdot\! \mathrm{Im}(\mathfrak{S}) + 0.335\cdot\! \mathrm{Im}(\mathfrak{T}) - \frac{m}{137\, p}\cdot\! A,
\label{NR_R}
 \end{eqnarray}
where $\mathfrak{S} \equiv (C_{S}+C'_{S})/C_{V}$, $\mathfrak{T} \equiv
(C_{T}+C'_{T})/C_{A}$ and $m$ is the electron mass.  The $R$
correlation coefficient vanishes in the lowest order SM calculations.
It becomes finite if final state interactions are included, $R_{FSI}\approx
-\frac{m}{137 p}\cdot A \approx 0.0006$, below the sensitivity
of this experiment. A larger value of $R$ would provide evidence
for the existence of exotic couplings, and a new source of time
reversal violation (TRV). Using Mott polarimetry, both
transverse components of the electron polarization can be measured
simultaneously: $\sigma_{T_{2}}$ perpendicular to the decay plane
defined by the neutron spin and electron momentum associated with $R$, and
$\sigma_{T_{1}}$ contained in the decay plane and associated with
$N$. The SM value of $N$ is finite and well within reach of this
experiment. Its determination provides an important test
of the experimental sensitivity.

\subsection{Experiment}
\label{nTRV:experiment}

The experiment was performed at the FUNSPIN beam line \cite{Zejma2005}
at the neutron source SINQ of the Paul Scherrer Institute, Villigen,
Switzerland. A detailed description of the design, operation and
performance of the Mott polarimeter can be found in \cite{Ban:2006qp}.
Only a short overview is presented here. The final result comprises
independent analyses of four data collection periods, featuring
different basic conditions such as beam polarization, Mott foil thickness
and acquired statistics.

\begin{figure}[t]
\centering
\includegraphics[width=0.6\textwidth,angle=270]{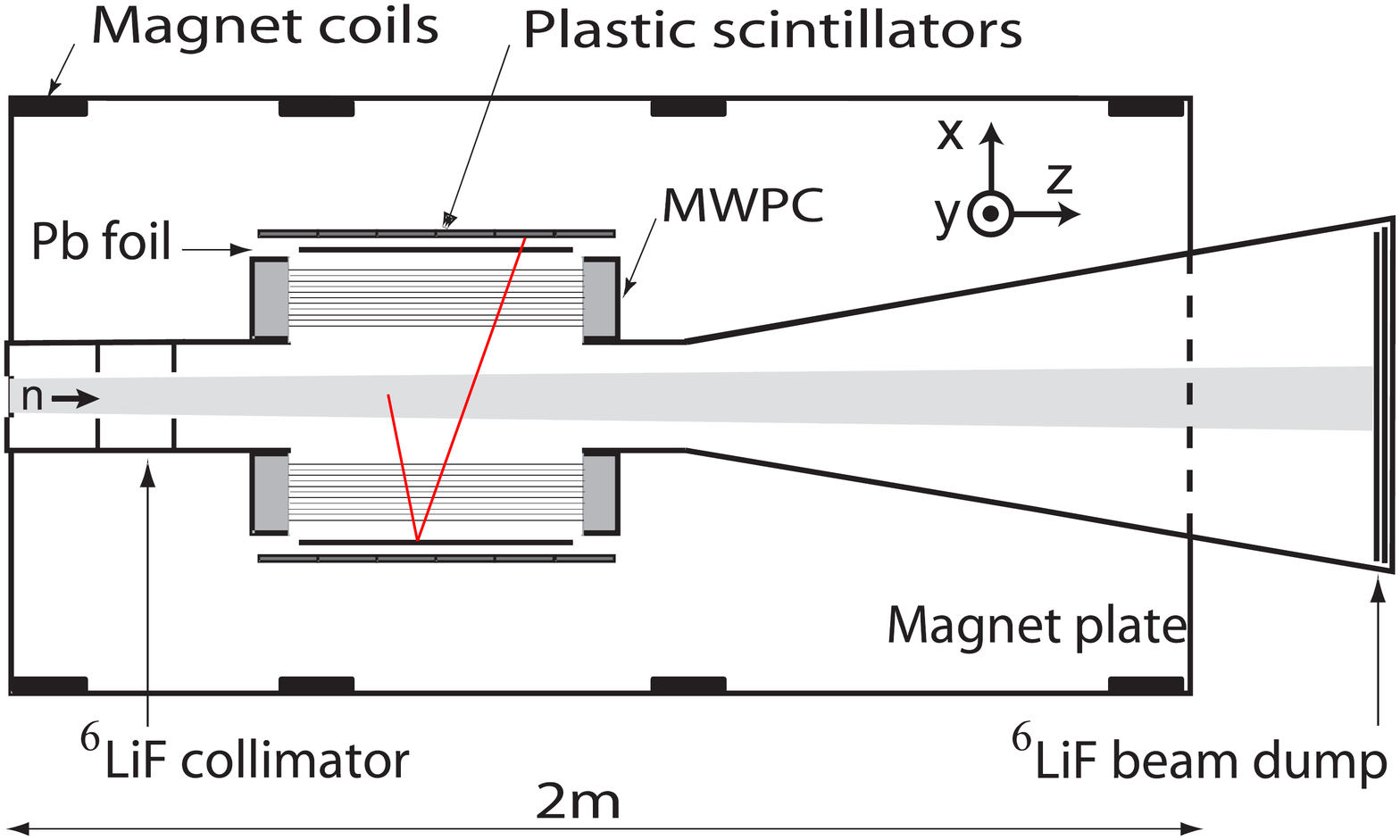}
\vspace*{-20pt}
\caption{\label{nTRV:fig1} Schematic top view of the experimental setup.  A~sample projection of an electron V-track event is indicated \cite{Kozela2009}.}
\vspace*{-7pt}
\end{figure}

The Mott polarimeter consisted of two identical modules, arranged
symmetrically on either side of the neutron beam
(\figref{nTRV:fig1}). The whole structure was mounted inside a
large-volume dipole magnet providing a homogeneous vertical spin-holding field of 0.5 mT within the beam fiducial volume. An RF-spin
flipper (not shown in \figref{nTRV:fig1}) was used to reverse the
orientation of the neutron beam polarization at regular time
intervals, typically every 16 s. Going outwards from the beam, each
module consisted of a multi-wire proportional chamber (MWPC) for
electron tracking, a removable Mott scatterer (1-2\,$\mu$m Pb layer
evaporated on a 2.5 $\mu$m thick mylar foil) and a scintillator
hodoscope to measure the electron energy.

A 1-cm-thick plastic scintillator, used for the electron energy
reconstruction, had a resolution of 33 keV at 500 keV.  The asymmetry
of the light signal collected at the ends of the scintillator slab
was used to determine the vertical hit position with a resolution of about
6 cm: the segmentation (10 cm) of the hodoscope in the horizontal
direction provided a crude estimate of the z-coordinate.  Matching
the information from the precise track reconstruction in the MWPC with
that from the scintillator hodoscope reduced background
and random coincidences considerably.

A 1.3-m-long multi-slit collimator defined the beam cross section to
4$\times$16 cm$^{2}$ at the entrance of the Mott polarimeter.
To minimize neutron scattering and capture, the entire beam
volume, from the collimator to the beam dump, was enclosed in a
chamber lined with $^{6}$Li polymer and filled with pure helium at
atmospheric pressure.  The total flux of the collimated beam was
typically about 10$^{10}$ neutrons/sec.  Thorough investigations of
the beam polarization performed in a dedicated experiment
\cite{Zejma2005} showed a substantial dependence on the position in
the beam fiducial volume.  The average beam polarization necessary for
the evaluation of the $N$- and $R$-correlation coefficients was
extracted from the observed decay asymmetry using the precisely known
\cite{Zyla2020} beta decay asymmetry parameter $A = -0.1196\pm0.0002$.
This approach automatically accounts for the proper integration over the
position-dependent beam density, its polarization and detector
acceptance.  For this purpose, single track events (only one
reconstructed track segment on the hit scintillator side) were
recorded using a dedicated prescaled trigger.  The main event trigger
was used to find V-track candidates: events with two reconstructed
segments on one side and one segment accompanied by a scintillator hit
on the opposite side, (see \figref{nTRV:fig1}).

The following asymmetries were analyzed to extract the beam polarization, $P$:
 \begin{equation}\label{nTRV:ASY1}
   \mathcal{E} \left(\beta, \gamma \right) =
\frac
{N^{+}\left(\beta, \gamma \right) -N^{-}\left(\beta, \gamma \right)}
{N^{+}\left(\beta, \gamma \right) +N^{-}\left(\beta, \gamma \right)}=
P \beta A cos(\gamma) ,
 \end{equation}
where $N^{\pm}$ are experimental, background-corrected counts of
single tracks sorted in 4 bins of the electron velocity $\beta$, and
15 bins of the electron emission angle $\gamma$ with respect to the
neutron polarization direction.  The sign in the superscripts reflects the
beam polarization direction.

\begin{figure}[t]
\centering
%\vspace*{-7pt}
\includegraphics[width=0.8\textwidth, trim=0 40pt 0 20pt, clip]{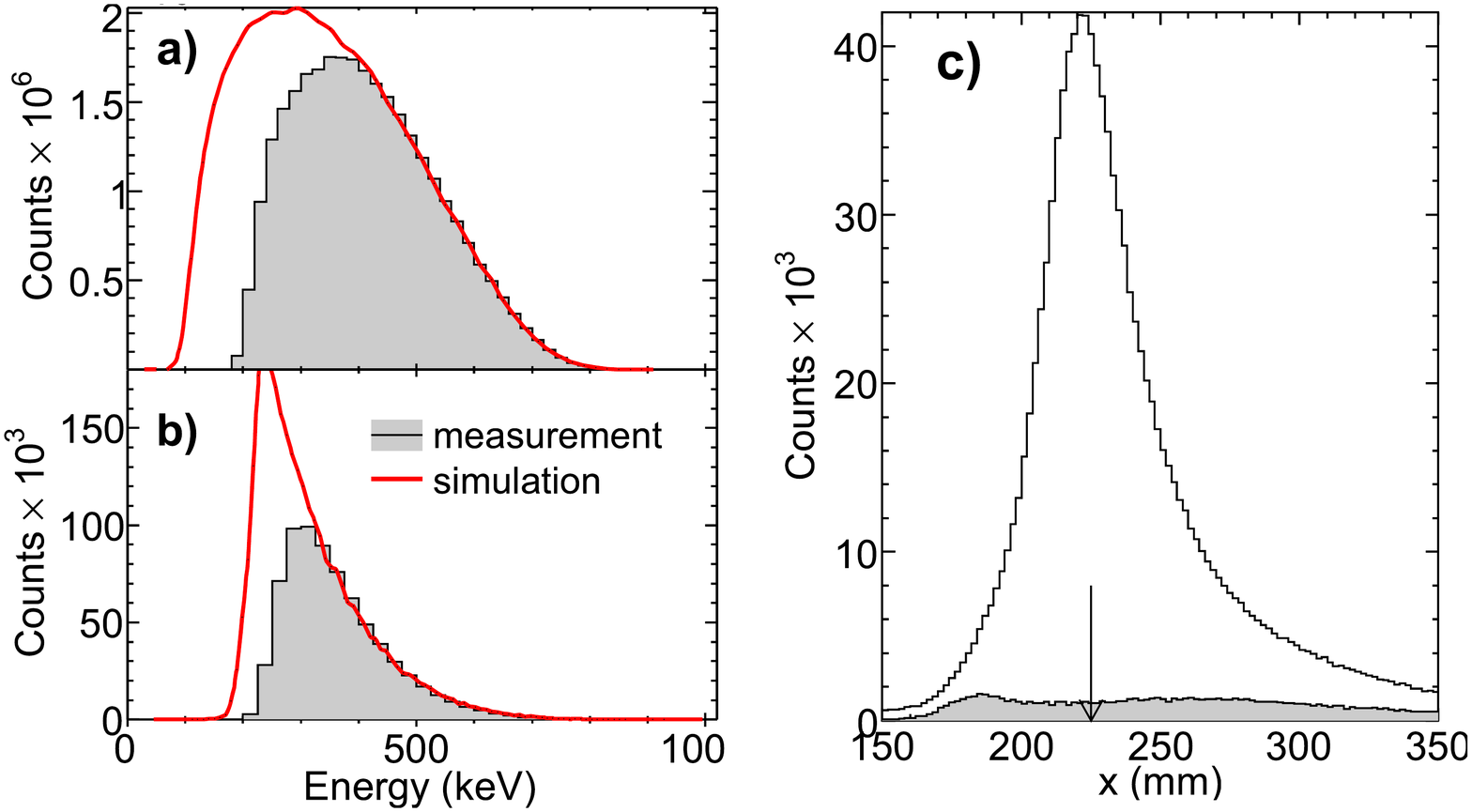}
%\vspace*{-7pt}
\caption{\label{nTRV:fig2} Background-corrected experimental energy
  distributions (shaded areas) of (a) the single-track and (b) V-track
  events compared with simulations. (c) Background contribution
  (shaded) to vertex $x$-coordinate distribution of V-track
  events. The arrow indicates the Mott foil position \cite{Kozela2009}.}
\vspace*{-5pt}
\end{figure}

A comparison between the measured and MC simulated energy spectra for
direct and Mott-scattered electrons is shown in \figref{nTRV:fig2}~a
and b, respectively.  Electronic thresholds are not included in the
simulation -- this is why the measured and simulated distributions do
not match at the low energy side.

Another set of asymmetries was used to extract the $N$ and $R$ correlation coefficients :
 \begin{equation}\label{nTRV:ASY3}
  \mathcal{A}\left(\alpha \right) =
\frac
{n^{+}\left(\alpha \right) - n^{-}\left(\alpha \right)}
{n^{+}\left(\alpha \right) + n^{-}\left(\alpha \right)} ,
 \end{equation}
where $n^{\pm}$ represent background-corrected experimental numbers of
counts of V-track events, sorted in 12 bins of $\alpha$,
the angle between electron scattering and neutron decay planes.  In
the case of V-track events, beside the background discussed
previously, events for which the scattering took place in the
surrounding of the Mott-target provide an additional source of
background.  \figref{nTRV:fig2}~c shows the distribution of the
reconstructed vertex positions in the $x$-direction for data collected
with and without the Mott foil.  The distribution clearly peaks at the
foil position.  This relatively broad distribution is a result of extrapolation of two electron track segments crossing at relatively small angle ($20^o-60^o$). Additionally, the electron straggling effects contribute to its broadening. The ``foil-out'' distribution has been scaled
appropriately by a factor deduced from the accumulated neutron beam.

It can be shown \cite{Ban:2006qp} that
 \begin{equation}\label{nTRV:ASY4}
   \mathcal{A}\left(\alpha \right) - P \bar{\beta} A  \mathcal{\bar{F}}(\alpha) =
	 P \bar{S}(\alpha) \left[ N  \mathcal{\bar{G}}(\alpha) +
	 R \bar{\beta}  \mathcal{\bar{H}}(\alpha) \right ] ,
 \end{equation}
where the kinematical factors $ \mathcal{\bar{F}}(\alpha)$, $
\mathcal{\bar{G}}(\alpha)$, and $ \mathcal{\bar{H}}(\alpha)$ represent
the average values of the quantities ${\bf \hat{J}\cdot\hat{p}}$,
${\bf \hat{J}}\cdot\mbox{\boldmath$ \hat{\sigma}$}$ and ${\bf
  \hat{J}\cdot\hat{p}}\times\!\mbox{\boldmath$ \hat{\sigma}$}$,
respectively, $\bar{S}$ is the effective analyzing power of the
electron Mott scattering, known in the literature as ``Sherman
function'', and the bar over a letter indicates event-by-event
averaging.  The term $P \bar{\beta} A \mathcal{\bar{F}}$ accounts for
the $\beta$-decay-asymmetry-induced nonuniform illumination of the
Mott foil.  Since the $\bar{\beta}$ and $ \mathcal{\bar{F}}$ are known
precisely from event-by-event averaging, the uncertainty of this term
is dominated by the error of the average beam polarization $P$.

Mean values of the effective analyzing powers as a function of
electron energy, scattering and incidence angles were calculated using
the Geant 4 simulation framework \cite{Agostinelli2003}, following
guidelines presented in \cite{Salvat2005,Khakoo2001}. This approach
accounts properly for the atomic structure, nuclear size
effects as well as the effects introduced by multiple scattering
in thick foils.

The systematic uncertainty is dominated by the effects introduced by
the background subtraction procedure, connected with the choice of the
geometrical cuts defining event classes ``from-beam'' and
``off-beam''.  To estimate this effect, the cuts were varied
in a range limited solely by the geometry of the apparatus.  Because the
radio--frequency of the spin flippers was a small source of noise
in the readout electronics, tiny spin-flipper-correlated dead time
variations were observed. The result was corrected for this effect.

The asymmetries as defined in \eqref{nTRV:ASY1} and \eqref{nTRV:ASY3}
have been calculated for events with energies above the neutron
$\beta$-decay end-point energy and for events originating outside of
the beam fiducial volume: they were found to be consistent with zero
within the statistical accuracy, which proves that the data were not biased
e.g. with a spin-flipper-related false asymmetry.

\begin{figure}[t]
\centering
%\vspace*{-7pt}
\includegraphics[width=0.8\textwidth, trim=0 70pt 0 20pt, clip]{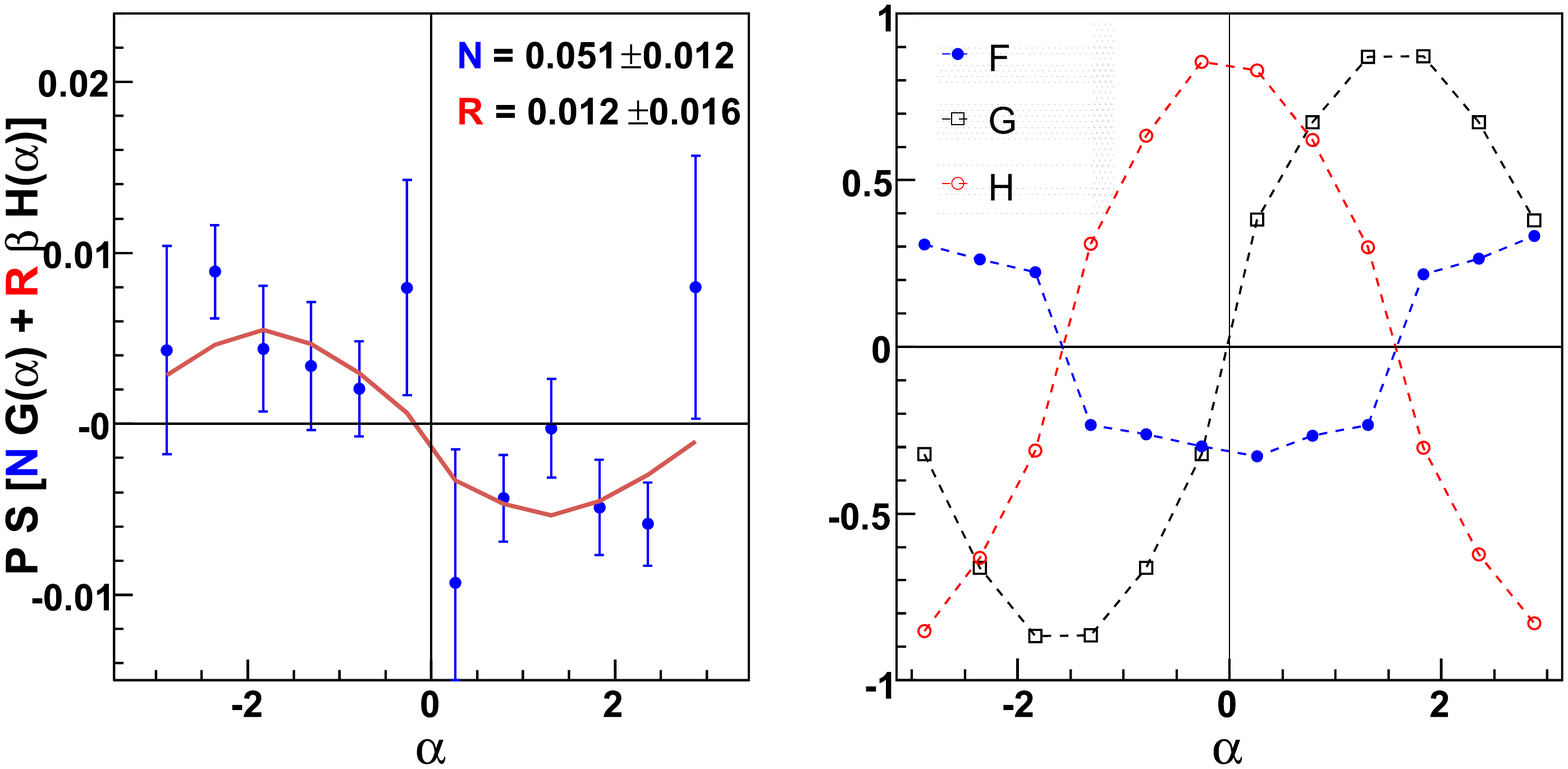}
%\vspace*{-10pt}
\caption{\label{nTRV:fig3}Left panel: experimental asymmetries $ \mathcal{A}$ corrected for the  $P \bar{\beta} A  \mathcal{\bar{F}}$ term for the 2007 data set as a function of $\alpha$ (defined in text).
The solid line illustrates a two-parameter ($N$, $R$)
least-square fit to the data.
The indicated errors are statistical.
Right panel: geometrical factors $ \bar{\mathcal{F}}(\alpha)$, $ \bar{\mathcal{G}}(\alpha)$ and $ \bar{\mathcal{H}}(\alpha)$ for the same data set \cite{Kozela2009}.}
\vspace*{-7pt}
\end{figure}

A fit of the experimental asymmetries $ \mathcal{A}$, corrected for the
 $P \bar{\beta} A  \mathcal{\bar{F}}$ term for the experimental data set of 2007 is shown in \figref{nTRV:fig3}.

From the approximate symmetry of the detector with respect to the transformation
$\alpha \rightarrow - \alpha$, it follows that  $ \bar{\beta}$, $\bar{S}$ and
the factors $ \mathcal{\bar{F}}$, $ \mathcal{\bar{H}}$ are all symmetric,
while $ \mathcal{\bar{G}}$ is an antisymmetric function of $\alpha$ (see
\figref{nTRV:fig3}).
This allows the extraction of the  $N$ coefficient from the expression \cite{Ban:2006qp}:
\begin{equation}\label{nTRV:SR1}
 N \approx  \frac{(r\!- \!1)}{(r\! +\!1)}\cdot
\frac{1-\frac{1}{2}(P\bar{\beta}A \bar{F})^2}{P \bar{S} \mathcal{\bar{G}}},\,\,\,
    r = \sqrt{\frac{n^{+}(\alpha) \, n^{-}(-\alpha )}
		{n^{-}(\alpha )\,  n^{+}(-\alpha )} }
\end{equation}
The advantage of this method is that the effect connected with the
term $P \bar{\beta} A \mathcal{\bar{F}}$ is suppressed by a factor of
about 60 compared to \eqref{nTRV:ASY4}.  The good agreement between
the $N$ values obtained in both ways enhances confidence in the
extracted $N$ and $R$ coefficient values.

The systematic uncertainties in the evaluation of the $R$ and
$N$ coefficients are dominated by effects introduced by the background
subtraction procedure and the choice of specific values of the cuts
that determine whether an individual event is attributed to
``signal'' or to ``background''. These effects were systematically
studied for all data sets. Additional calibration
measurements were performed to determine the Mott-target mass
distribution \cite{Kozela2010} that can influence the electron
depolarization leading to increased uncertainty of the effective
Sherman function. A detailed description of the data analysis process
can be found in \cite{Kozela2009,Kozela2012} together with the
final result comprising all available experimental data.
\begin{eqnarray}\label{nTRV:FinalResult}
 R &=& 0.004 \pm 0.012_{\mathrm{stat}} \pm 0.005_{\mathrm{syst}}, \\
 N &=& 0.067 \pm 0.022_{\mathrm{stat}} \pm 0.004_{\mathrm{syst}}.
%\nonumber
\end{eqnarray}
This was the first determination of the $N$ correlation coefficient in $\beta$-decay.

\begin{figure}[t]
\centering
\includegraphics[width=0.9\textwidth]{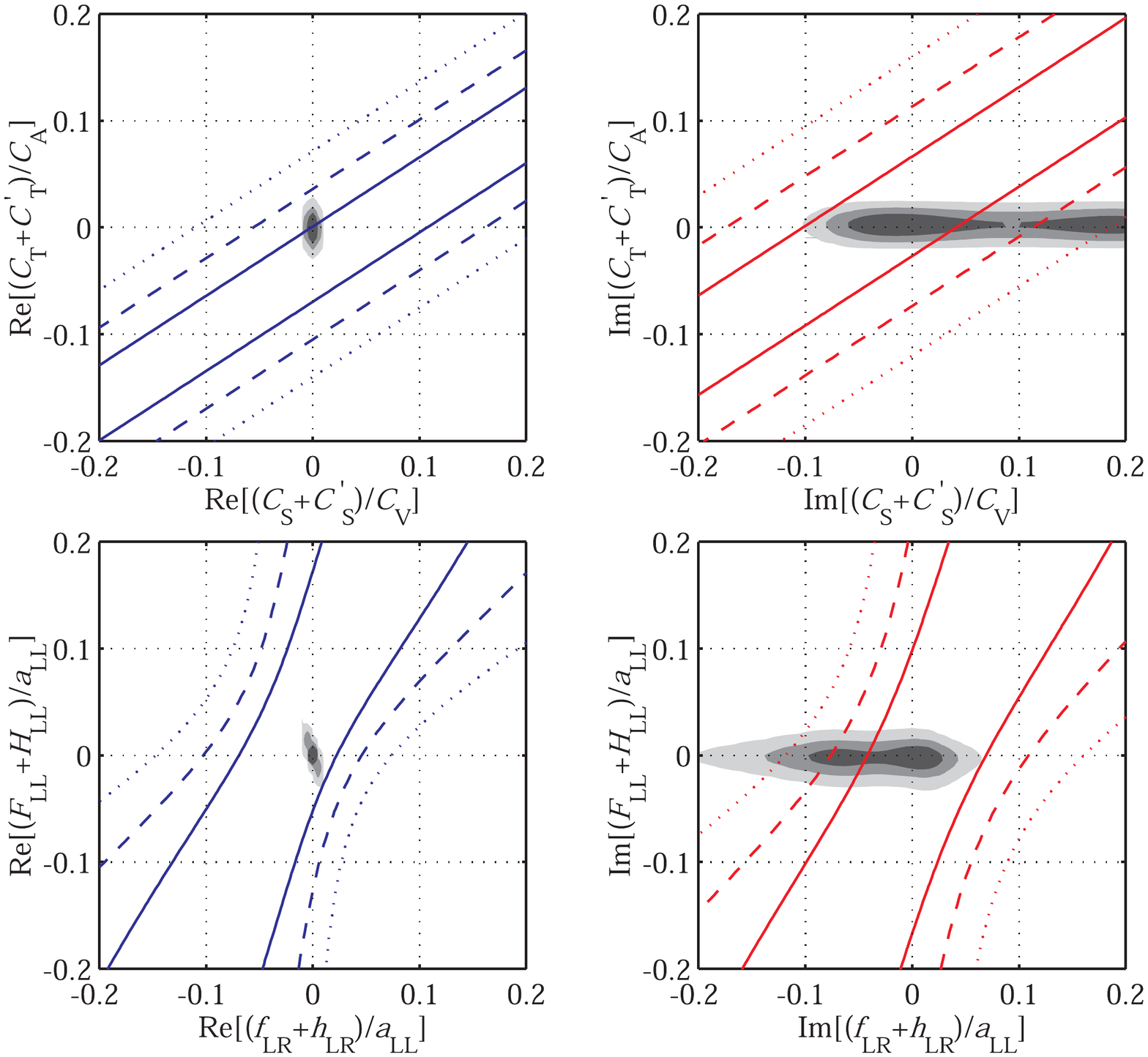}
\caption{\label{nTRV:fig4}Experimental bounds on the scalar vs. tensor
  normalized couplings (upper) and leptoquark exchange helicity
  projection amplitudes (lower panels) published in \cite{Kozela2009}.
  The gray areas represent the available to date empirical
  information as listed in \cite{Severijns2006}, while the lines
  represent the limits resulting from the present experiment. Solid,
  dashed and dotted lines correspond to 1-, 2- and 3- sigma confidence
  levels, respectively, in analogy to decreasing intensity of the grey
  areas.}
\vspace*{-12pt}
\end{figure}

In \figref{nTRV:fig4} the new results are included in exclusion
plots containing all experimental information available from nuclear
and neutron beta decays as surveyed in \cite{Severijns2006}. The
upper plots contain the normalized scalar and tensor coupling
constants $\mathfrak{S}$ and $\mathfrak{T}$, while the lower plots
correspond to the helicity projection amplitudes in the leptoquark
exchange model, as defined in \cite{Herczeg2001}.  Although the
achieved accuracy does not improve the already strong constraints on
the real part of the couplings (left panels), the result is
consistent with the existing data and increases
confidence in the validity of the extraction of $R$.  For the
imaginary part (right panels), the new experimental value of the $R$
coefficient significantly constrains scalar couplings beyond the
limits from all previous measurements.  The result is consistent
with the SM.

\subsection{Outlook -- the BRAND project}
\label{nTRV:outlook}
The successful determination of two transverse components of the
polarization of electrons emitted in neutron decay in a pioneering and
nearly optimal experiment led to the following conclusions: (i) it
seems quite possible to decrease the systematic uncertainty by an
order of magnitude using existing techniques, (ii) the transverse electron polarization can be studied in a more systematic way by correlating it with
the electron momentum, the neutron spin, and also with the
recoil proton momentum by constructing
larger and higher acceptance detecting systems like e.g. proposed by
\cite{Sromicki2000} and operating with the highest intensity
polarized cold neutron beam available. In this way, one can study seven correlation
coefficients: $H$, $L$, $N$, $R$, $S$, $U$ and $V$ where five of them
($H$, $L$, $S$, $U$, $V$) have never been experimentally studied:
\begin{align}\label{nTRV:EQN_JTW1}
&\omega(E_e,\Omega_e,\Omega_{\bar{\nu}}) \;\propto\; 1 \; +\nonumber \\
 &a\,\frac{\textbf{p}_e\cdot\textbf{p}_{\bar{\nu}}}{E_eE_{\bar{\nu}}} + b\,\frac{m_e}{E_e}
+ \frac{\langle\textbf{J}\rangle}{J}\cdot\left[ A\,\frac{\textbf{p}_e}{E_e} + B\,\frac{\textbf{p}_{\bar{\nu}}}{E_{\bar{\nu}}} + D\,\frac{\textbf{p}_e\times\textbf{p}_{\bar{\nu}}}{E_eE_{\bar{\nu}}} \right] \;+ \nonumber \\
& \boldsymbol\sigma_\perp\cdot\left[H\,\frac{\textbf{p}_{\bar{\nu}}}{E_{\bar{\nu}}} + L\,\frac{\textbf{p}_e\times\textbf{p}_{\bar{\nu}}}{E_eE_{\bar{\nu}}} +  N\,\frac{\langle\textbf{J}\rangle}{J} + R\,\frac{\langle\textbf{J}\rangle\times\textbf{p}_e}{J\,E_e}  \right. + \; \nonumber \\
&\;\;\;\;\;\;\;\left. S\,\frac{\langle\textbf{J}\rangle}{J}\frac{\textbf{p}_e\cdot\textbf{p}_{\bar{\nu}}}{E_eE_{\bar{\nu}}} + U\,\textbf{p}_{\bar{\nu}}\frac{\langle\textbf{J}\rangle\cdot\textbf{p}_e}{J\,E_eE_{\bar{\nu}}} + V\,\frac{\textbf{p}_{\bar{\nu}}\times\langle\textbf{J}\rangle}{J\,E_{\bar{\nu}}} \right] ,
\end{align}
where $\boldsymbol\sigma_\perp$ represents a unit vector perpendicular
to the electron momentum $\textbf{p}_e$ and
$J=|\textbf{J}|$. $\textbf{p}_{\bar{\nu}}$ and $E_{\bar{\nu}}$ are the
antineutrino momentum and energy, respectively.

The coefficients relating the transverse electron polarization to
$\textbf{p}_e$, $\textbf{p}_{\bar{\nu}}$ and $\textbf{J}$ have several
interesting features. They vanish for the SM weak interaction, and reveal
the variable size of the electromagnetic contributions. For $H$ and $N$, the electromagnetic
contributions are of the order of 0.06, which can be used for an
internal sensitivity check of the Mott polarimeter. Finally, the dependence on the real and imaginary parts of the scalar and
tensor couplings alternates exclusively from one correlation
coefficient to another with varying sensitivity. This feature allows
a complete set of constraints to be determined from the neutron decay
alone.

The idea of implementing such a complex measurement was proposed
in \cite{Bodek2011}. An updated version of the measurement can be
found in \cite{Bodek2018}. Presently, the first test run devoted to
the verification of the applied detectors and techniques has been
completed on the PF1B cold neutron beam at the Laue Langevin Institute
in Grenoble, France (ILL).

\begin{figure}[t]
  \centering
  \includegraphics[width=0.8\textwidth]{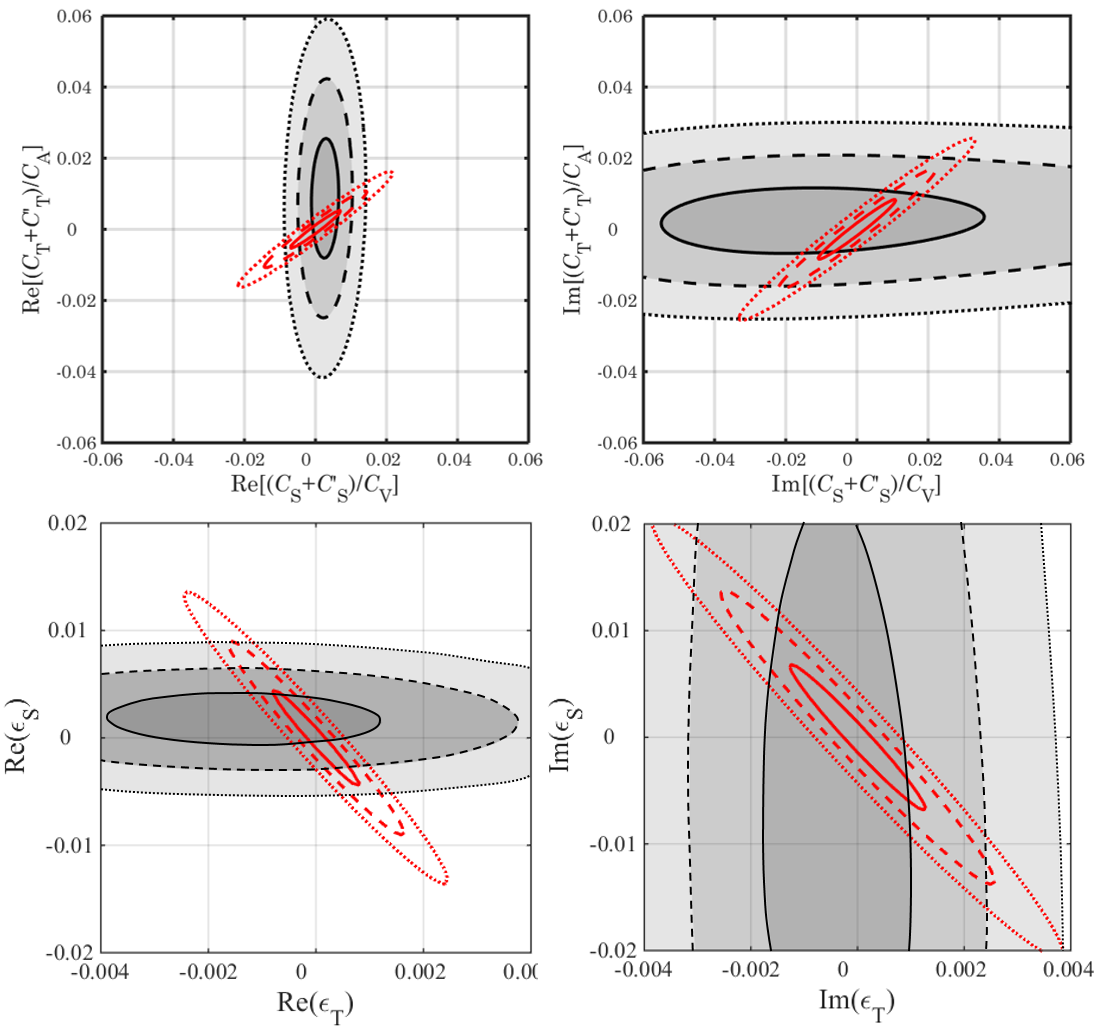}
  \caption{\label{nTRV:fig5} Experimental bounds on the scalar
    vs. tensor couplings $\mathfrak{S}$, $\mathfrak{T}$ from
    \eqref{nTRV:NR_N} (upper panels) and translated to EFT parameters
    $\epsilon_S$, $\epsilon_T$ (lower panels) published in \cite{Bodek2018}. The gray areas
    represent the information deduced from available
    experiments as listed in \cite{Gonzalez2019}, while the red lines represent the limits resulting
    from the correlation coefficients $H$, $L$, $N$, $R$, $S$, $U$ and
    $V$ measured with the anticipated accuracy of $5 \times
    10^{-4}$. Solid, dashed and dotted lines correspond to 1-, 2- and
    3-$\sigma$ confidence levels, respectively, in analogy to
    decreasing intensity of the grey areas.}
\end{figure}

\subsection{EFT parameterization}
\label{nTRV:EFT}
To bridge the classical $\beta$-decay formalism with
high-energy physics and permit sensitivity comparison of low-energy
charged-current observables with measurements carried out at
high-energy colliders, the model-independent effective field theory
(EFT) framework is employed. The effective nucleon-level couplings
$C_i$, $C'_i$ ($i\in[V,A,S,T]$) can be generally expressed as
combinations of the quark-level parameters $\epsilon_i$,
$\tilde{\epsilon}_i$ ($i\in[L,R,S,T]$) \cite{Naviliat2013}. The
imaginary parts of the scalar and tensor couplings parameterize
CP-violating contributions. The high energy BSM physics process that
can be compared with $\beta$-decay experiments is the cross section
for electrons and missing transverse energy (MET) in $pp\rightarrow
e\bar{\nu}+MET+\ldots$ channel since it has the same underlying
partonic process as in $\beta$-decay ($\bar{u}d\rightarrow
e\bar{\nu}$). With the anticipated accuracy of about $5\times
10^{-4}$ for the transverse electron polarization related correlation
coefficients in the BRAND experiment one would obtain significantly tighter bounds on the real
and imaginary parts of scalar and tensor coupling constants and,
consequently, on $\epsilon_S$ and $\epsilon_T$ as shown in
\figref{nTRV:fig5}. It should be noted that such limits would be
competitive to those extracted from the analysis of 20 fb$^{-1}$ CMS
collaboration data collected at 8 TeV
\cite{Chatrchyan2012,Khachatryan2013} and even to the planned
measurements at 14~TeV.

\subsection*{Acknowledgments}
This work has been supported in part by The National
Science Centre, Poland, under the grant No. 2018/29/B/ST2/02505.

% =========================================================================

\bibliography{nTRV}

\nolinenumbers

\end{document}